\documentclass[aps,prb,showpacs,twocolumn,superscriptaddress]{revtex4}
\usepackage{graphicx}
\usepackage{bm}
\usepackage{amsmath}

\def\be{\begin{equation}}       \def\ee{\end{equation}}
\def\bea{\begin{eqnarray}}      \def\eea{\end{eqnarray}}

\begin{document}

\title{Tuning phase transition between quantum 
spin Hall and ordinary insulating phases}
\author{Shuichi Murakami}
\email[Electronic address: ]{murakami@stat.phys.titech.ac.jp}
\affiliation{Department of Physics, Tokyo Institute of Technology,
2-12-1 Ookayama, Meguro-ku, Tokyo 152-8551, Japan}
\author{Satoshi Iso}
\affiliation{Institute of Particle and Nuclear Studies,
High Energy Accelerator Research Organization (KEK)
Tsukuba 305-0801, Japan}
\author{Yshai Avishai}
\affiliation{Department of Physics,
Ben-Gurion University , Beer-Sheva 84105 Israel}
\affiliation{RTRA project, LPS (Orsay) and CEA (Saclay), France}
\author{Masaru Onoda}
\affiliation{Correlated Electron Research Center (CERC), 
National Institute of Advanced
Industrial Science and Technology (AIST), Tsukuba Central 4, Tsukuba 
305-8562, Japan}
\affiliation{
CREST, Japan Science and Technology Corporation (JST), Saitama, 332-0012, 
Japan}
\author{Naoto Nagaosa}
\affiliation{Department of Applied Physics, University of Tokyo,
Hongo, Bunkyo-ku, Tokyo 113-8656, Japan}
\affiliation{
CREST, Japan Science and Technology Corporation (JST), Saitama, 332-0012, 
Japan}
\affiliation{Correlated Electron Research Center (CERC), 
National Institute of Advanced
Industrial Science and Technology (AIST), Tsukuba Central 4, Tsukuba 
305-8562, Japan}

\begin{abstract}
An effective theory is constructed for analyzing a generic 
phase transition between the quantum spin Hall and the 
insulator phases.  Occurrence of degeneracies due to
closing of the gap at the transition are carefully elucidated. 
For systems without inversion symmetry 
the gap-closing occurs at $\pm \vec{k}_{0}(\neq \vec{G}/2)$ 
while for 
systems with inversion symmetry, the gap can close only at wave-numbers
$\vec{k}=\vec{G}/2$, where $\vec{G}$ is a reciprocal lattice vector.
In both cases, following a unitary transformation which mixes spins,
the system is represented by two
decoupled effective theories of massive two-component fermions 
having masses of opposite signs.
Existence 
of gapless helical modes at a domain wall between the two phases
directly follows from this formalism. This theory provides an elementary and comprehensive phenomenology of the quantum spin Hall system.
\end{abstract}
\pacs{
73.43.-f,       
72.25.Dc,	
73.43.Nq 	
85.75.-d        
}

\maketitle
\section{Introduction}
The intrinsic spin Hall effect (SHE) \cite{Murakami03a,Sinova04} driven by the relativistic spin-orbit interaction and the associated Berry curvature of the Bloch wavefunctions attracts considerable attention both theoretically and experimentally. In conducting materials such as doped GaAs, the external electric field produces transport current and dissipation, even though the spin current transverse to it is dissipationless. Therefore it is highly desirable to design systems showing the SHE without dissipation. Spin Hall insulator (SHI) has been proposed for this purpose by some of the present authors, and its candidate materials are HgTe, PbTe, $\alpha$-Sn and so forth \cite{Murakami04c}. These band insulators are predicted to show finite spin Hall conductivity $\sigma^s_H$, which is not quantized and depends on parameters in the model Hamiltonian. Later it has been realized that gapless edge modes in semi-infinite systems do not exist in generic cases. These two features, i.e., the non-quantized $\sigma^s_H$ and the absence of gapless edge modes, are closely related to 
the absence of conserved spin current in the presence of spin-orbit interaction, i.e., there is no U(1) gauge symmetry for spin current. Therefore it was difficult to distinguish between the SHI and usual insulators.

Recently,  Kane and Mele proposed a model for time-reversal (T-)
invariant systems \cite{Kane05a,Kane05b}, which manifests a finite SHE 
and demonstrated its distinction from an ordinary insulator due to the topological nature of its ground state. The pertinent $Z_2$ topology is represented by an integer $\Delta$ defined for Bloch wavefunctions in the bulk, whose parity distinguishes the relevant phases. Physically, $\Delta$ is identical with the number of pairs of helical edge modes. In a system with $\Delta=odd$, referred to as quantum spin Hall (QSH) system \cite{Kane05a,Kane05b,Onoda05a,Bernevig05a,Qi05,Sheng06,Fu06a,Fu06b,Fu06c,Murakami06b,Onoda07}, the odd number of pairs of helical edge modes is robust against weak nonmagnetic disorder and interactions \cite{Wu05,Xu05}. When $\Delta=even,$ 
gapless edge modes 
can hybridize each other and a gap will open even at the edge. The system is then referred to as spin Hall insulator (SHI). Transitions between phases with $\Delta=even$ (SHI) and $\Delta=odd$ (QSH) occur only when the gap is closed by tuning parameters of the model. Constructing a theory for analyzing these transitions is a challenge of paramount interest.

In this paper, we develop an effective continuum theory for phase transitions between QSH and SHI systems in 2D 
and discuss (i) classification of the possible types of transition, (ii) existence of gapless helical edge modes, and (iii) the change of the $Z_2$ topological number at the transition. The basic idea is that effective continuum theory focusing on the vicinity of the gap-closing points at the transition can be constructed even though characterization of each phase requires information over the whole first Brillouin zone. Namely, the {\it change} across the phase boundary is much easier to elucidate, and the relation between the topological number and the helical edge modes is rather transparent. 
This work concerns local features in $\vec{k}$ space, and 
is complementary to Refs.~\onlinecite{Kane05b,Moore07}, which treats
global topological
structure in $\vec{k}$ space.
We ignore interaction and disorder effect in this paper, since the robustness of the system is inferred from the topological stability.

\section{Gap closing at the phase transition}
Since the phase transition necessarily accompanies closing of the 
gap \cite{Kane05b,Sheng06, Fukui07}, 
we commence with an analysis of generic gap-closing in 
a two-dimensional (2D) gapped spin-$1/2$ T-symmetric system with spin-orbit interaction.
A Hamiltonian matrix for Bloch wavefunctions can be written in 
a block form,
\begin{equation}
H(\vec{k})=\left(\begin{array}{cc}
h_{\uparrow\uparrow}(\vec{k})& h_{\uparrow\downarrow}(\vec{k})\\
h_{\downarrow\uparrow}(\vec{k})& h_{\downarrow\downarrow}(\vec{k})
\end{array}\right).\label{eq:Hamiltonian}
\end{equation}
The dimension of the matrix 
$h_{\sigma\sigma'}(\vec{k})$ depends on systems considered;
nevertheless,
in order to describe the phase transition,
it is sufficient to 
restrict the dimension of the matrices $h_{\sigma \sigma'}(\vec{k})$ 
to be one or two,
as we will 
see later.
To investigate the topological order of the Hamiltonian, its spectrum 
is assumed to have a gap, within which 
the Fermi energy lies. The
T-symmetry is represented by the operator, $\Theta=i \sigma_{y} K$ 
($\sigma_{x,y,z}$ are Pauli matrices and $K$ stands for complex conjugation). 
For $H(\vec{k})$ it implies,
\begin{equation}
H(\vec{k})=\sigma_y  H^{T}(-\vec{k})\sigma_y,
\label{time-reversal}\end{equation}
or, equivalently, 
$h_{\uparrow\uparrow}(\vec{k})
=h_{\downarrow\downarrow}^{T}(-\vec{k})$ and
$h_{\uparrow\downarrow}(\vec{k})
=-h_{\uparrow\downarrow}^{T}(-\vec{k})$.
This, in turn, results in a degeneracy between states at $\vec{k}$ and 
 $-\vec{k}$, forming Kramers pairs.

Tuning some parameters in the Hamiltonian 
may drive a transition, where the gap closes 
and degeneracies between the valence and the conduction bands occur
 at certain wavevectors 
$\vec{k}=(k_x,k_y)$. To pursue the phase transition, 
we will focus on  {\it ``generic'' } gap-closing achieved by tuning
 a {\it single} parameter $m$. 
(For mere convenience, the critical value of $m$ for which a generic gap-closing occurs 
is chosen as $m=0$.) 
Non-generic 
gap-closing achieved by tuning {\it several} parameters are excluded in our analysis.
This is because such kind of gap-closing can be circumvented by small perturbation, meaning 
that it cannot be associated with a phase transition.
As we show below, 
generic gap-closing are
classified into two cases shown 
schematically in Fig.~\ref{fig:degeneracy} (a)(b) (depending on symmetry under parity).
We note that while we have not made any assumption on the $Z_2$ topological number, 
both cases (a) and (b) turn out to encode
quantum phase transitions between the QSH and the SHI phases.
Among the known models describing this kind of phase transition,
the Kane-Mele model on the honeycomb lattice \cite{Kane05b} falls
 within class (a) while the HgTe quantum well model \cite{Bernevig06f} belongs to class (b).

The QSH-SHI phase transitions pertaining to Fig.~\ref{fig:degeneracy} (a)(b) are not so trivial as it 
might look. In general, energy levels repel each other, thereby the 
valence and the conduction bands do not touch when the number of tuned 
parameters is not large enough. The number of tuned parameters to 
achieve degeneracy, called the codimension, is sensitive to the symmetry 
and the dimension of the system considered. For example, in three dimensions
the gap-closing of the type (a) in Fig.~\ref{fig:degeneracy} does not occur
\cite{Murakami07c}.

Consider now a spatial inversion (I-)symmetry which plays an important role 
beside T-symmetry. The former requires the relation 
$\varepsilon_{n \alpha}(\vec{k}) = \varepsilon_{n \alpha}(-\vec{k})$, 
while the latter implies
$\varepsilon_{n \alpha}(\vec{k}) = \varepsilon_{n{\bar \alpha}}(-\vec{k})$,
where $\varepsilon_{n \alpha}(\vec{k})$ is the energy of 
 band $n$ with pseudospin $\alpha$, and ${\bar \alpha}$
is the pseudospin opposite to $\alpha$ 
\cite{note-pseudospin}.
If both symmetries are respected, 
$\varepsilon_{n \alpha}(\vec{k}) = \varepsilon_{n {\bar \alpha}}(\vec{k})$
and there is a Kramers double degeneracy at each $\vec{k}$.
If I-symmetry is broken, double degeneracy occurs at points 
 $\vec{k} = - \vec{k} + \vec{G}$, i.e. $\vec{k}=\vec{k}_{i}\equiv
\vec{G}/2$, where $\vec{G}$ is a reciprocal lattice vector;
no double degeneracy occurs at other points
(unless an additional symmetry is present).
The I-asymmetric and I-symmetric cases
are therefore considered separately below. 

\subsection{Inversion asymmetric systems}
 In this case, 
when $\vec{k} \neq - \vec{k} + \vec{G}$,
each band is {\it non-degenerate}, and 
the gap
between two bands can close
at some points $\vec{k}$. 
At the gap-closing point, one valence band and one conduction band
become degenerate. 
The codimension is three 
\cite{vonNeumann29}. 
This codimension three is equal to the number of parameters involved, 
that is, $k_x$, $k_y$ and $m$. Thus the gap can close at some $\vec{k}$
when the parameter $m$ is tuned to a critical value.

On the other hand, when $\vec{k} = - \vec{k} + \vec{G}$, 
the band
is doubly degenerate, and the codimension 
is five \cite{Avron88,Avron89}, exceeding  the number of
tunable parameters which is one (that is, $m$). 
Thus, generic gap-closing
cannot occur at $\vec{k}=\vec{G}/2$ \cite{note-1}.

\begin{figure}
\includegraphics[scale=0.37]{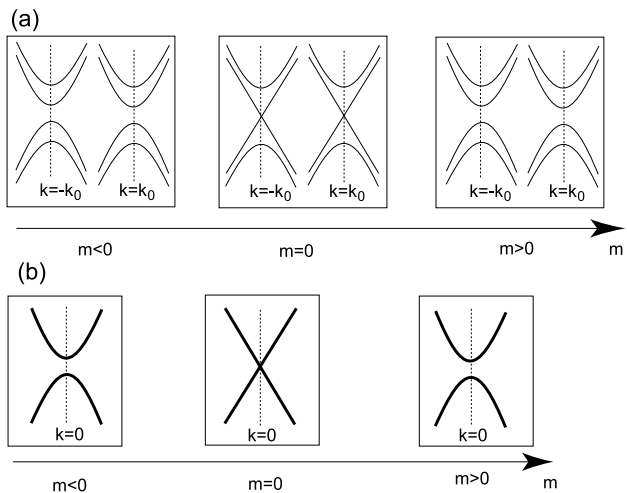}
\caption{Generic gap-closing for (a) inversion-asymmetric and 
(b) inversion-symmetric cases. In case (b) all the states are
doubly degenerate by Kramers theorem.}
\label{fig:degeneracy}\end{figure}

We thus focus on the case $\vec{k}\neq \vec{G}/2$.
Near the gap-closing point $\vec{k}=\vec{k}_{0}(\neq \vec{G}/2)$,
the system's Hamiltonian corresponds to massive two-component fermion, 
and can be expressed 
as ${\cal H}=m\sigma_z 
+(k_x-k_{0x})\sigma_x+(k_y-k_{0y})\sigma_y$
(after unitary and scale transformations). T-symmetry requires that
the gap closes simultaneously at $\vec{k}_0$ and 
$-\vec{k}_0$ as depicted in 
Fig.~\ref{fig:degeneracy} (a), and that the masses at $\vec{k}=\pm 
\vec{k}_{0}$ have 
opposite signs.
In the honeycomb-lattice model for QSH \cite{Kane05a,Kane05b}
the gap closes at the $K,K'$ points;
Hence it reduces to the present scheme without I-symmetry.

\subsection{Inversion symmetric systems}
 In this case, the energies 
are doubly degenerate
at each $\vec{k}$. At the phase transition, 
the gap between the two doubly-degenerate bands closes. 
Hence, we consider 4$\times$4 Hamiltonian 
matrix 
$H(\vec{k})$ ($h_{\alpha \beta}(\vec{k}$) in 
Eq.~(\ref{eq:Hamiltonian}) are 2$\times$2). 
The I-symmetry is imposed as
\begin{equation}
H(-\vec{k})=PH(\vec{k})P^{-1}, \ u(-\vec{k})=Pu(\vec{k}),
\label{eq:H-inversion}
\end{equation}
where $P$ is a unitary matrix independent of $\vec{k}$, 
and $u(\vec{k})$ is the periodic part 
of the Bloch wavefunction: $\varphi_{\vec{k}}(\vec{r})
=u(\vec{k})e^{i\vec{k}\cdot\vec{r}}$.
As the inversion does not flip spin, this 
unitary matrix $P$ should be block-diagonal in spin space:
\begin{eqnarray}
P=\left(\begin{array}{cc}
P_{\uparrow}&\\&P_{\downarrow}
\end{array}
\right).\label{eq:P}
\end{eqnarray}
In fact, all cases reduce to a case
$P_{\uparrow}=P_{\downarrow}=\text{diag}(\eta_{a},\ \eta_{b})$ with 
$\eta_{a}=\pm 1$, $\eta_{b}=\pm 1$, 
as can be shown directly by applying a unitary transformation.
 $\eta_a$ and $\eta_b$ represent the
parity eigenvalues of the atomic orbitals.

Occurrence of gap closing turns out to be different for the cases
(i) $\eta_a=\eta_b$ and 
(ii) $\eta_a=-\eta_b$.
The case (i) $\eta_{a}=\eta_{b}=\pm 1$ is realized when 
the atomic orbitals $a$, $b$ 
have the same parity, such as two $s$-like orbitals or two $p$-like 
orbitals. The ensuing constraints on the Hamiltonian are:
$h_{\uparrow\uparrow}=h_{\downarrow\downarrow}^{T}$ is an even function 
of $\vec{k}$, and $h_{\uparrow\downarrow}=h_{\downarrow\uparrow}^{\dagger}$ is
an antisymmetric matrix, even function in $\vec{k}$.
In an explicit form, the generic Hamiltonian 
becomes
\begin{equation}
{H}(\vec{k})=E_{0}(\vec{k})+\sum_{i=1}^{5}a_{i}(\vec{k})
\Gamma_{i}\end{equation}
where $a_{i}$'s and $E_{0}$ are real even functions of $\vec{k}$.
$\Gamma_{i}$ are $4\times 4$ matrices given by
$\Gamma_{1}=1 \otimes
\tau_{x}$, $\Gamma_{2}=\sigma_{z}\otimes\tau_{y}$, 
$\Gamma_{3}=1 \otimes\tau_{z}$, 
$\Gamma_{4}=\sigma_{y}\otimes\tau_{y}$, and  
$\Gamma_{5}=\sigma_{x}\otimes\tau_{y}$, where 
$\sigma_{i}$ and $\tau_{i}$ are Pauli matrices acting
on spin and orbital spaces, respectively.
The Eigenenergies are given by 
$E_{0}\pm\sqrt{\sum_{i=1}^{5}a_{i}^{2}}$. 
The gap closes when $a_{i}(\vec{k})=0$ for $i=1,\cdots,5$.
It means that the codimension 
is five, the same as in the case at $\vec{k}=\vec{G}/2$ without I-symmetry. 
Thus there are no solutions of $k_x$, $k_y$ and $m$ which satisfy these
five relations. 
The gap never closes in this case.

Next we consider the case $\eta_{a}=-\eta_{b}=\pm 1$, i.e. $P=\eta_{a}
\tau_{z}
=\pm\tau_{z}$,
where the two constituent atomic orbitals have different parity.
The Hamiltonian reads,
\begin{equation}
{H}(\vec{k})=a_{0}(\vec{k})+a_{5}(\vec{k})\Gamma'_{5}+\sum_{i=1}^{4}
b^{(i)}(\vec{k})
\Gamma'_{i}
\end{equation}
where $a_0(\vec{k})$ and $a_{5}(\vec{k})$ are even functions of $\vec{k}$, and 
$b^{(i)}(\vec{k})$ are odd functions of $\vec{k}$. Here
$\Gamma'_{i}$  are $4\times 4$ matrices given by  
$\Gamma'_{1}=\sigma_{z}\otimes\tau_{x}$,
$\Gamma'_{2}=1 \otimes\tau_{y}$,
$\Gamma'_{3}=\sigma_{x}\otimes\tau_{x}$,
$\Gamma'_{4}=\sigma_{y}\otimes\tau_{x}$,
and $\Gamma'_{5}=1 \otimes\tau_{z}$.
In this case the gap closes only when five equations $a_{5}(\vec{k})=0$,
$b^{(i)}(\vec{k})=0$ are satisfied. 
For a generic point $\vec{k}$ with $\vec{k}\neq \vec{G}/2$, 
these five equations cannot be satisfied 
simultaneously, through a change of a single parameter \cite{note-2}.
On the other hand, at the high-symmetry points $\vec{k}=\vec{k}_{i}
=\vec{G}/2$,  the situation is different. At these points
the odd functions $b^{(i)}(\vec{k})$ vanish identically, 
and one has only to tune
$a_{5}(\vec{k})$ to be zero. Thus, the gap closes by fine-tuning a single parameter.
To be more specific, we take $\vec{k}=0$ as an example, and
write down the Hamiltonian explicitly. Extension to other $\vec{k}=
\vec{G}/2$ points is straightforward.
The Hamiltonian is expanded to linear order in $\vec{k}$ as
\begin{equation}
{H}(\vec{k})\sim E_{0}+m\Gamma'_{5}+\sum_{i=1}^{4}
\left(\vec{\beta}^{(i)}\cdot\vec{k}\right)
\Gamma'_{i},
\end{equation}
where $E_0$ and $m$ are constants, and 
$\vec{\beta}^{(i)}$ $(i=1,\cdots,4)$ are two-dimensional real constant vectors.
Further simplification is obtained after judicious 
unitary transformations.
The Hamiltonian finally acquires the block-diagonal form,
\begin{equation}
{H}(\vec{k})=E_{0}+\left(
\begin{array}{cccc}
m&z_{-}&&\\
z_{+}&-m&&\\
&&m&-z_{+}\\
&&-z_{-}&-m
\end{array}\right).\label{eq:case-b}\end{equation}
where $z_{\pm}=b_{1}k_{x}+b_{3}k_{y}\pm ib_{2}k_{y}$ with real constants
$b_1$, $b_2$ and 
$b_3$.
Note that in materials with high crystallographic symmetry (e.g tetragonal),
one has $b_{1}=b_{2}$ and $b_{3}=0$, leading to $z_{\pm}\propto 
k_{x}\pm ik_{y}$.
We have thus demonstrated a feature:
{\it The Hamiltonian of a generic system with spin-orbit coupling 
obeying T- and I- symmetries 
decouples, after an appropriate choice of basis, into a pair of 
Hamiltonians describing two-component fermions
 with opposite sign of the 
corresponding mass terms.}
(Such decoupling is expected in the special case where $s_z$ is a good 
quantum number, since the system 
describes  two copies of a quantum Hall system). Experimental consequences 
are immediate as the Hamiltonian (\ref{eq:case-b})
is equivalent to the one  suggested for the HgTe quantum well 
in Ref.~\onlinecite{Bernevig06f} (based on phenomenological arguments).
Gap closing at $\vec{k}=0$ when 
the parameter $m$ is tuned to zero is now obvious, since the eigenenergies are
$E=E_{0}\pm \sqrt{m^{2}+z_{+}z_{-}}$. 
The inversion matrix in this basis 
is written as 
$\eta_a\otimes \tau_z=\eta_a\mathrm{diag}(1,-1,1,-1)$.

Summing up to this point, we discussed generic types of gap closing
in time-reversal invariant 
systems, achieved by tuning a single parameter. 
Taking I-symmetry into consideration, 
two types of gap-closing scenarios have been found: (a) 
simultaneous gap closing at $\vec{k}=\pm\vec{k}_{0}\neq\vec{G}/2$ occur
in systems without I-symmetry, and (b) gap closing between
two Kramers-degenerate bands (i.e. four bands) at $\vec{k}=\vec{G}/2$ occur in systems with 
I-symmetry 
(see Fig.~\ref{fig:degeneracy}). 
Due to the level repulsion, the gap-closing by tuning a single parameter
occurs only in limited cases.
We will see in the subsequent discussion that these cases of gap-closing 
exactly coincides with the phase transition between 
the QSH and the insulating phases. In this sense our theory characterizes the 
QSH phase from the local features in $\vec{k}$ space.

\noindent
\subsection{ Change of the $Z_2$ topological number at the gap-closing point}
Now we focus on the change of $Z_2$ topological number at the gap closing,
assuming that except for the gap closing ($m=0$) the bands are fully gapped.
Hence, both phases at $m>0$ and $m<0$ 
are band insulators, and have well-defined $Z_2$ topological 
numbers. 
For the I-symmetric case (a), the homotopy characterization in 
Ref.~\onlinecite{Moore07} 
is applicable; for the lower band at the critical value $m=0$, 
there is one vortex at $\vec{k}_{0}$ and one antivortex at 
$-\vec{k}_{0}$. Thus, when the parameter $m$ is tuned across $m=0$, 
the Chern number for the whole contracted surface \cite{Moore07}
changes by one. Thus, the $Z_2$ topological 
numbers are different by one for the $m>0$ and  the $m<0$ sides.
One of the phases is the QSH phase, while the other is the SHI.
For the I-symmetric case (b), Fu and Kane \cite{Fu06c}
developed a simple 
method to calculate the $Z_2$ topological number $\Delta$ as 
\begin{equation}
(-1)^{\Delta}=\prod_{i=1}^{4}\prod_{m=1}^{N}\xi_{2m}(\vec{k}_{i}),
\end{equation}
where $\vec{k}_i$ are the four high-symmetry points satisfying $\vec{k} =
\vec{G}/2$,
$\xi_{2m}(k_i)$ is the parity eigenvalue at each of these
points, and 
$N$ is the number of Kramers pairs below $E_F$.
In the present case the gap at $\vec{k}=0$ collapses  when $m=0$. Hence only 
 the parity eigenvalue at $\vec{k}=0$ can change at the phase transition. 
Since the inversion matrix is given by ${P}=
\eta_{a}\otimes\tau_z=\eta_{a}\sigma_{0}\otimes\tau_z$, 
the parity eigenvalues are $-\eta_{a}(=\eta_{b})$ and $+\eta_{a}$ for 
the lower-band 
states at $m>0$ and $m<0$, respectively. Hence, the parity 
eigenvalue changes sign, and the $Z_2$ topological number $\Delta$ 
changes by one. 
Thus, on the two sides of the band touching, $m>0$ and $m<0$, 
one of the phases is the QSH phase, while the other one is the simple 
insulator (SHI) phase.

\section{Helical edge states}
Let us regard the usual insulating phase as our vacuum, so that the 
domain wall between the QSH phase and the insulating phase
is the edge of the sample. Such a domain wall is described by
a spatially dependent mass parameter $m(x)$ satisfying 
 $m(\pm \infty)= \pm m_{0}$, i.e., 
\begin{equation}
m=\left\{\begin{array}{l}
m_{0}\ \ : x\gg 0\\
-m_{0}\ \ : x\ll 0.
\end{array}\right.
\label{eq:DW}
\end{equation}
We do not specify the detail of the crossover between $m_{0}$ and
$-m_{0}$, because it is not important for the subsequent discussions.
For Fig.~\ref{fig:degeneracy}(a), one can 
consider the Weyl fermions at $\vec{k}=\pm \vec{k}_{0}$ separately.
Masses of these Weyl fermions change sign at $m=0$; hence they yield 
the edge states localized at the domain wall, as explained in 
Ref.~\onlinecite{Niemi86}.
Because the Weyl fermions at $\vec{k}=\pm \vec{k}_{0}$ are 
related to each other by time-reversal symmetry, 
the two edge states form a Kramers pair.

For Fig.~\ref{fig:degeneracy}(b) with a domain wall 
(\ref{eq:DW}), we follow the discussion in 
Ref.~\onlinecite{Niemi86} to 
show that such domain wall between phases with different $Z_2$ 
topological number possesses one Kramers pair of edge states at the 
boundary.
In this case we consider 
\begin{widetext}
\begin{equation}
\tilde{H}(k_{y})=E_{0}+\left(
\begin{array}{cccc}
m&-ib_1\partial_x +(b_3 -ib_2)k_y&&\\
-ib_1\partial_x +(b_3 +ib_2)k_y&-m&&\\
&&m&ib_1\partial_x -(b_3 +ib_2)k_y\\
&&ib_1\partial_x -(b_3 -ib_2)k_y&-m
\end{array}\right).
\label{eq:case-b-DM}
\end{equation}
To calculate the eigenstates it is convenient to perform unitary
transformation as
\begin{equation}
H'(k_y)=Q^{\dagger}\tilde{H}(\vec{k})Q=
E_{0}+\left(
\begin{array}{cccc}
b_2 k_y& m-b_1\partial_x&&\\
m+b_1\partial_x &-b_2 k_y&&\\
&&-b_2 k_y& m-b_1\partial_x\\
&&m+b_1\partial_x&b_2 k_y
\end{array}\right),\label{eq:case-b-DM-u}\end{equation}
\end{widetext}
where 
\begin{equation}
Q=e^{-ib_3 k_{y}x/b_{1}}\left(
\begin{array}{cccc}
1&1&&\\
i&-i&&\\
&&-i&-i\\
&&-1&1
\end{array}
\right).\end{equation}
We omit the term $E_0$ henceforth, since it does not affect the 
subsequent discussions. 
The eigenvalue problem reads as $H'(k_y)u_{k_{y}}(x)=E(k_y)u_{k_y}(x)$.

Because (\ref{eq:case-b-DM-u}) is block-diagonal, we first solve 
the eigenvalue problem for the first two components of $u_{k_{y}}$, i.e 
we put $u_{k_{y}}=(u_1,\ u_2,\ 0,\ 0)^{t}$. We get
\begin{eqnarray}
&&(E-b_2 k_y)u_1=Du_{2},\label{eq:Du2} \\
&&(E+b_2 k_y)u_2=D^{\dagger}u_{1}, \label{eq:Du1} 
\end{eqnarray}
where $D=m-b_{1}\frac{\partial}{\partial x}$, 
$D^{\dagger}=m+b_{1}\frac{\partial}{\partial x}$.
They yield eigenequations for $u_{1}$ and $u_{2}$, respectively:
\begin{eqnarray}
&&DD^{\dagger}u_{1}=(E^{2}-b_{2}^{2}k_{y}^{2})u_{1},\label{eq:DDdag}\\
&&D^{\dagger}Du_{2}=(E^{2}-b_{2}^{2}k_{y}^{2})u_{2},\label{eq:DdagD}
\end{eqnarray}
When we solve Eq.~(\ref{eq:DDdag}) 
for $u_{1}$, one can calculate $u_{2}$
from (\ref{eq:Du1}). As (\ref{eq:DDdag}) is invariant under
$E\rightarrow -E$, it 
seems that the solutions for $E$ and $-E$ are always obtained simultaneously,
namely there is a spectral symmetry between $E\leftrightarrow -E$. 
Nevertheless, it does not apply if $E=-b_2 k_y$, where (\ref{eq:Du1})
cannot be solved for $u_{2}$. A similar situation occurs for 
$E=b_2 k_y$. Thus 
exceptions at $E=\pm b_{2}k_{y}$ occur in the following way.
For $u_{1}(\neq 0)$ which satisfies $D^{\dagger}u_{1}=0$,
we get $E=b_{2}k_{y}$ and $u_2=0$ from Eqs.~(\ref{eq:Du2}) and
(\ref{eq:Du1}), whereas there is no 
solution with $E=-b_{2}k_{y}$. In the same token, for 
$u_{2}$ which satisfies $Du_{2}=0$, we get 
 $E=-b_{2}k_{y}$ from (\ref{eq:Du2}), whereas there is no 
solution with $E=b_{2}k_{y}$. 
Hence the spectral asymmetry is related to 
the kernels for $D$ and $D^{\dagger}$.
For simplicity we take $b_1>0$ henceforth, 
while the other case of $b_1<0$ can 
be studied in a similar way. For the domain wall (\ref{eq:DW}), 
the solution of $D^{\dagger}u_{1}=0$ gives 
\begin{equation}
u_{1}\propto
\exp\left(-b_{1}^{-1}\int^{x}m(s)ds\right)
\end{equation} and $E=b_{2}k_{y}$, 
while $Du_{2}=0$ has no normalizable solution. 
Thus the energy dispersion in $k_{y}$ direction has a branch $E=b_{2}k_{y}$,
which crosses the Fermi energy $E\sim 0$. This state is gapless, 
localized near $x=0$.

\begin{figure}[h]
\includegraphics[scale=0.6]{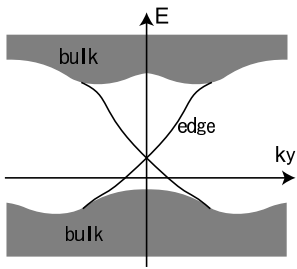}
\caption{Schematic dispersion curves for 
the model (\ref{eq:case-b-DM}). Kramers theorem  guarantees that the crossing 
of two edge states occur at $k_y=0$. }
\label{fig:dispersion}\end{figure}
So far we have solved the eigenequation for the first two components.
The lower two components of the wavefunction $u$ is obtained from above by 
time-reversal operation ($\Theta u(k_{y})= i\sigma_{2}u(-k_{y})^{*}$ ).
Therefore, the above-mentioned edge state with $E=b_2 k_y$ has a Kramers
partner with $E=-b_{2} k_{y}$. The whole dispersion is shown in 
Fig.~\ref{fig:dispersion}.
Thus we have shown that 
the Kramers pair of edge states exists 
at the boundary between the two phases.
They cross at $k_y=0$, as is guaranteed by the Kramers theorem.

\section{Conclusions}
A general framework is established for 
classifying phase transitions between the quantum spin Hall and the 
insulator phases.  For inversion-asymmetric 
systems, the phase transition accompanies a gap closing at $\vec{k}=\pm 
\vec{k}_{0}$ which is not at the high-symmetry points. For 
inversion-symmetric systems, the gap closes only
at $\vec{k}=\vec{G}/2$ where $\vec{G}$ is a reciprocal vector.
All the known models  exhibiting phase transition between 
the two phases are special cases of this general classification framework.

\begin{acknowledgments}
This research is supported in part 
by Grant-in-Aid and NAREGI Nanoscience Project 
from the Ministry of Education,
Culture, Sports, Science and Technology of Japan.  
Y.~A. is supported by the Invitation Fellowship for Research in Japan 
through the Japanese Society for the Promotion of Science.

\end{acknowledgments}


\begin{thebibliography}{99}
\bibitem{Murakami03a}
S. Murakami, N. Nagaosa, and S.-C. Zhang, Science \textbf{301}, 1348 (2003).
\bibitem{Sinova04}
J. Sinova, D. Culcer, Q. Niu, N. A. Sinitsyn, T. Jungwirth, and A. H. MacDonald, Phys. Rev. Lett. \textbf{92}, 126603 (2004).
\bibitem{Murakami04c}
S. Murakami, N. Nagaosa, and S.-C. Zhang, Phys. Rev. Lett. \textbf{93}, 
156804 (2004).
\bibitem{Kane05a}
C. L. Kane and E. J. Mele, Phys. Rev. Lett.
\textbf{95}, 146802 (2005).
\bibitem{Kane05b}
C. L. Kane and E. J. Mele, Phys. Rev. Lett.
\textbf{95}, 226801 (2005).
\bibitem{Onoda05a}
M. Onoda and N. Nagaosa, Phys. Rev. Lett. \textbf{95}, 106601 (2005).
\bibitem{Bernevig05a}
B. A. Bernevig and S.-C. Zhang,
Phys. Rev. Lett. \textbf{96}, 106802 (2006).
\bibitem{Qi05}
X.-L. Qi, Y.-S. Wu and S.-C. Zhang,
Phys. Rev. B\textbf{74},085308 (2006).
\bibitem{Sheng06}
D. N. Sheng, Z. Y. Weng, L. Sheng, and F. D. M. Haldane, Phys. Rev. Lett. \textbf{97}, 036808 (2006).
\bibitem{Fukui07}
T. Fukui and Y. Hatsugai, Phys. Rev. B\textbf{75}, 121403(R) (2007).
\bibitem{Fu06a}
L. Fu and C. L. Kane, Phys. Rev. B\textbf{74}, 195312 (2006).
\bibitem{Fu06b}
L. Fu, C. L. Kane, and E. J. Mele, Phys. Rev. Lett. \textbf{98},
106803 (2007).
\bibitem{Fu06c}
L. Fu and C. L. Kane, 
Phys. Rev. B\textbf{76}, 045302 (2007).
\bibitem{Murakami06b}
S. Murakami, Phys. Rev. Lett. \textbf{97}, 236805 (2006).
\bibitem{Onoda07}
M. Onoda, Y. Avishai, N. Nagaosa, Phys. Rev. Lett. \textbf{98}, 076802 (2007).
\bibitem{Wu05}
C. Wu, B. A. Bernevig, and S.-C. Zhang, Phys. Rev. Lett. \textbf{96}, 106401
(2006).
\bibitem{Xu05}
C. Xu and J. E. Moore,
Phys. Rev. B\textbf{73}, 045322 (2006).
\bibitem{Bernevig06f}
B. A. Bernevig, T. L. Hughes, S.-C. Zhang,
Science \textbf{314}, 1757 (2006).
\bibitem{Murakami07c}
S. Murakami, New J. Phys. {\bf 9}, 356 (2007).
\bibitem{note-pseudospin}
The pseudospin $\alpha$, which labels the  eigenstates, is 
different from the spin in the presence of the spin-orbit coupling.
\bibitem{vonNeumann29}
V. J. von Neumann and E. Wigner, Physik. Zeitschr. \textbf{30}, 467 (1929).
\bibitem{Avron88}
J. E. Avron, L. Sadun, J. Segert, and B. Simon, 
Phys. Rev. Lett. \textbf{61}, 1329 (1988).
\bibitem{Avron89}
J. E. Avron, L. Sadun, J. Segert, and B. Simon, 
Commun. Math. Phys. \textbf{124}, 595 (1989).
\bibitem{note-1}
By considering all possible 2D magnetic 
point groups, we found some exceptional models with degeneracy at $\vec{k}=\vec{G}/2$ , even without I-symmetry. These cases result from a rather high symmetry of the system. It is expected that by including some perturbation 
to lower the symmetry 
the Hamiltonian will reduce to the case (a) in Fig.~\ref{fig:degeneracy}.
\bibitem{note-2}
In some exceptional cases with high point-group symmetry,
the gap between 
two doubly degenerate bands can close even when 
$\vec{k}\neq \vec{G}/2$. In 2D this occurs  for the 
magnetic point-group symmetries 
$\bar{3}'$, $ \bar{3}'m$, $\bar{3}'m'$, $6'/m$, $6'/mm'm$,
which are in trigonal system or hexagonal systems.
It is expected that 
the closing of the gap in these cases does not correspond to phase transition 
between the QSH and the SHI, as some perturbation can 
circumvent this degeneracy. 
\bibitem{Moore07}
J. E. Moore and L. Balents, Phys. Rev. B \textbf{75}, 121306(R) (2007) .
\bibitem{Niemi86}
A similar problem has been discussed in the context of fractional fermion number. See A. J. Niemi and G. W. Semenoff, Phys. Rep. \textbf{135}, 99 (1986) for a review.
A concrete example  is the midgap state associated with dimerized 
soliton in polyacetylene as discussed in  W.~P.~ Su, J.~R.~Schrieffer, and A. J. Heeger, Phys. Rev. Lett. \textbf{42}, 1698 (1979).
\end{thebibliography}
\end{document}